\documentclass[prl,aps,twocolumn,showpacs,amsmath,amssymb,amsfonts,floatfix,superscriptaddress]{revtex4}
\usepackage{graphicx}
\usepackage{dcolumn}
\usepackage{bm}
\usepackage[hypertex]{hyperref}
\usepackage{latexsym}
\usepackage{float}
\usepackage{supertabular}
\usepackage{longtable}

\begin{document}
\title{Metastability and anomalous fixation in evolutionary games on scale-free networks}
\author{Michael Assaf\footnote[1]{The authors contributed equally to this work.}}
\affiliation{Loomis Laboratory of Physics, Department of Physics 1110 West Green Street, Urbana, Illinois 61801, U.S.A}
\author{Mauro Mobilia\footnotemark[1]}
\affiliation{Department of Applied Mathematics, School of Mathematics, University of Leeds, Leeds LS2 9JT, U.K.} %
\begin{abstract}
We study the influence of complex graphs on the metastability and fixation properties
of a set of evolutionary processes. In the framework of evolutionary game theory,
where the fitness and selection are frequency-dependent and vary with the population composition, we analyze the dynamics of
snowdrift games (characterized by a metastable coexistence state) on  scale-free networks. Using an effective diffusion theory in the weak selection limit, we
demonstrate how the scale-free structure  affects the system's metastable
state and leads to {\it anomalous fixation}. In particular, we analytically and numerically show that the probability and mean time of fixation are characterized by stretched exponential behaviors with exponents
depending on the network's degree distribution.
\end{abstract}
\pacs{05.40.-a, 02.50.Ey, 87.23.Kg, 64.60.aq}

\maketitle
The evolutionary dynamics of systems where successful traits spread at the expense of others
is naturally modeled in the framework of evolutionary game theory (EGT)~\cite{EGT,Nowak}.
In EGT, each species reproductive potential (fitness) varies with the  population's composition
and changes  continuously in time. The selection is therefore ``frequency-dependent'' and the dynamics
is traditionally studied in terms of differential equations~\cite{EGT,Nowak,freqdepsel}.
Evolutionary dynamics is known to be affected by
demographic noise and by the population's spatial arrangement
~\cite{Spatialcoop,weaksel}, and
 is often characterized by the central notion of fixation.
This refers to the possibility that a ``mutant type'' takes over~\cite{Kimura}, and one is particularly
interested in the {\it fixation probability} -- the probability that a given trait invades an
 entire population, and in the {\it mean fixation time} (MFT) -- the mean time for this event to occur.
In contrast to what happens in spatially-homogeneous (well-mixed) populations,  the  spatial arrangement of
individuals can give rise to very different scenarios~\cite{Spatialcoop,EGT}.
Evolutionary dynamics  on networks~\cite{Nets} provides a general and unifying framework  to describe the dynamics of
both well-mixed and spatially-structured populations~\cite{EGTgraphs,EGTgraphs-bis}. In spite of its importance,
fixation  of evolutionary processes on networks has been mostly studied in idealized situations,
\textit{e.g.} for two-state systems under a constant weak selective bias~\cite{EGTgraphs,EGTgraphs-bis,VotNets3,VotNets,VotNets2}. In these works, it has been shown that the update rules and the network structure effectively renormalize the population size and  thereby affect the fixation properties.
Furthermore,  some properties of evolutionary games have been studied on scale-free networks
by numerical simulations, see \textit{e.g.}~\cite{Pacheco},
and on regular graphs with  mean field and perturbative treatments~\cite{det-pertub-net}.
The models  of Refs.~\cite{EGTgraphs,VotNets3,VotNets,VotNets2}
are of great interest but  do {\it not} provide a general description of evolutionary dynamics on graphs.
In particular, these references consider constant fitness and selection pressure, and thus can
 \textit{not} describe systems possessing a long-lived {\it metastable} coexistence state prior to fixation~\cite{AM,MA10}.

In this Letter we study metastability, which may arise as a consequence of frequency-dependent selection~\cite{freqdepsel}, and fixation on a class of scale-free networks in the EGT framework. To the best of our knowledge, such an analytical study has not been conducted before.
For concreteness, we investigate ``snowdrift games'' (SGs, see below)~\cite{EGT,MA10} that are the
paradigmatic EGT models exhibiting metastability (see \cite{Gore09} for their experimental relevance). Our findings are
also directly relevant to various fields,
 \textit{e.g.}   to population genetics~\cite{FixNet} and to the dynamics of epidemic outbreaks, for which a long-lived endemic state is often an intrinsic characteristic~\cite{AM,Durrett,Nasell}.

For well-mixed populations (complete graphs) the fixation properties of  SGs
typically exhibit an exponential dependence on the population size, see {\it e.g.}~\cite{MA10}.
Our central result is the demonstration that  evolutionary dynamics on scale-free
networks can lead to {\it anomalous} fixation and metastability characterized by a stretched exponential dependence
on the population size, in stark contrast with their non-spatial counterparts.
In the same vein as in \cite{VotNets}, the analytical description is based on an effective diffusion theory derived from an
individual-based formulation of the dynamics.

{\it The model.} We consider a network comprising $N$ nodes, each of which is either occupied by an individual of type
$\textsf{C}$ (cooperator) or by a $\textsf{D}$-individual (defector).
The occupancy of the node $i$ is encoded by the random variable $\eta_i$, with
$\eta_i=1$ if the node $i$  is occupied by a $\textsf{C}$ and $\eta_i=0$ otherwise.
The state of the system is thus described by  $\{ {\bm \eta}\}=\{\eta_i\}^N$
and the density of cooperators present in the system is $\rho\equiv \sum_{i=1}^N \eta_i/N$.
The network is specified by its adjacency
matrix ${\bm A}=[A_{ij}]$, whose elements are $1$ if the nodes $ij$ are connected and
$0$ otherwise. The network is also characterized by its degree distribution $n_k=N_k/N$, where $N_k$ is the number of
nodes of degree $k$. EGT is traditionally concerned with large and homogeneous populations (i.e. $N\to \infty$
and $A_{ij}=1,\forall ij$) whose mean field dynamics is described by the celebrated replicator equation~\cite{EGT,Nowak}:
$(d/dt)\rho(t)=\rho(t)(1-\rho(t))[\Pi^{C}(\rho(t))-\Pi^{D}(\rho(t))]$, where
$\Pi^{C/D}(\rho(t))$ are the cooperator/defector average payoffs  derived from
the game's payoff matrix.
For a generic two-strategy cooperation dilemma, the payoff of  $\textsf{C}$  against another
 $\textsf{C}$ is denoted $a$ and that of  $\textsf{D}$ playing against $\textsf{D}$ is   $d$. When $\textsf{C}$ plays against $\textsf{D}$
the former gets payoff $b$ and the latter gets $c$~\cite{EGT}.
Here we focus on SGs, for which  $c>a$ and $b>d$. SGs are characterized by a stable interior fixed point
$\rho_*=(d-b)/(a-b-c+d)$ and unstable absorbing states $\rho=0$ (all-$\textsf{D}$) and $\rho=1$ (all-$\textsf{C}$).
For a finite population size ($N< \infty$) the role of fluctuations is important and  $\rho_*$ becomes a
{\it metastable} state whose decay time on complete graphs ($A_{ij}=1,\forall ij$) grows exponentially with $N$~\cite{MA10}.

In a spatial setting, the interactions are among
 nearest-neighbor  individuals and the species
payoffs are  defined locally:  $\textsf{C}$ and $\textsf{D}$ players at  node $i$
interacting with a neighbor at node $j$ respectively
receive payoffs    $\Pi_{ij}^C=a\eta_j+b(1-\eta_j)$ and $\Pi_{ij}^D=c\eta_j+d(1-\eta_j)$.
In the spirit of the Moran model (in the weak selection limit)~\cite{Nowak,weaksel,Kimura},
each species  local reproductive potential, or fitness, is given by the difference of $\Pi_{ij}^{C/D}$
relative to the population mean payoff $\bar{\Pi}_{ij}(t)$.
Here, we make the mean-field-like choice
$\bar{\Pi}_{ij}(t)=\rho(t) \Pi_{ij}^C +(1-\rho(t))\Pi_{ij}^D$
to include what arguably is the simplest mechanism ensuring the formation of metastability.
It is customary to introduce a selection strength $s>0$
in the definition of the fitness to unravel the  interplay between random fluctuations and selection~\cite{Nowak,weaksel,Kimura}. Here, the  fitnesses of $\textsf{C}/\textsf{D}$
at node $i$ interacting with a neighbor at node $j$ are
\begin{eqnarray}
f_{ij}^C=1+s[\Pi_{ij}^C-\bar{\Pi}_{ij}] \;\; \text{and}\;\;
 f_{ij}^D=1+s[\Pi_{ij}^D-\bar{\Pi}_{ij}].
\end{eqnarray}
These expressions comprise a baseline contribution (set to $1$) and a selection term proportional to the relative
 payoffs. Moreover, we  consider a system evolving according to the so-called ``link dynamics''
(LD)~\cite{VotNets3,VotNets}: a link is randomly selected at each time step and if it connects a  $\textsf{C}\textsf{D}$ pair,
one of the neighbors is randomly selected for reproduction with a rate proportional to its fitness, while the other is
replaced by the offspring.
While various types of update rules are possible~\cite{dynamics},
we here use the LD to highlight the combined effects of the topology and frequency-dependent selection:
here, in stark contrast to the LD
in the constant selection/fitness scenario~\cite{VotNets},
we show that the fixation properties strongly depend on
the network's heterogeneity.
Moreover, we have checked that  our conclusion
is robust and  holds for various other
update rules leading to metastability~\cite{inprep}.

The evolution of the population's composition
is described in terms of $\{\rho_k\}$, where $\rho_k=\sum_{i}' \eta_i/N_k$ is
the average number of cooperators  on all nodes of degree $k$ (the prime denotes summation over degree $k$ nodes), i.e $\rho_k$ is
the  subgraph density of $\textsf{C}$'s  on nodes of degree $k$.
 Quantities  necessary for our analysis are the $m^{th}$ moment of the degree distribution, $\mu_m\equiv \sum_k k^m n_k=\sum_i k_i^m /N$, where $k_i$ denotes the degree of node $i$, and the degree-weighted density of cooperators $\omega\equiv\sum_k (k/\mu_1) n_k \rho_k$.

{\it Effective diffusion theory.}
To implement the evolutionary dynamics, we
introduce $\Psi_{ij}=(1- \eta_i) \eta_j f_{ji}^C$ and $\Psi_{ji}=(1-\eta_j)\eta_i f_{ji}^D$,
where $\eta_i (1- \eta_j)$ is non-zero only when the nodes $ij$ are occupied by a
$\textsf{C}\textsf{D}$ pair.
In  the LD, the probability to select the neighbor $j$
of node $i$ for an update
is $A_{ij}/(N\mu_1)$ and the transition  $\eta_i \to 1-\eta_i$ hence
occurs with probability $ \sum_j \frac{A_{ij}}{N\mu_1} \left[\Psi_{ij} +\Psi_{ji}\right]$~\cite{VotNets}.
The subgraph density  $\rho_k$
changes by $\pm \delta \rho_k=\pm 1/N_{k}$ according to a birth-death process~\cite{Gardiner} defined by the
transition rates $T^{+}(\rho_k)=\sum_{i}'\sum_j A_{ij}
\Psi_{ij}/(N\mu_1)$ and $T^{-}(\rho_k)=\sum_{i}'\sum_j A_{ij}
\Psi_{ji}/(N\mu_1)$, respectively.
For our analytical treatment, we focused on degree-heterogeneous networks
with degree-uncorrelated nodes, as in Molloy-Reed networks (MRN)~\cite{Molloy},
yielding $A_{ij}=k_ik_j/(N\mu_1)$. Our numerical simulations
 were performed using the ``redirection algorithm'' that generates degree-correlated
scale-free networks~\cite{redir}. Yet, it has been shown that the dynamics on the latter is close to that on MRN~\cite{VotNets}.
With  $\sum_{i}'N^{-1}=n_k\rho_k$, the transition rates  become
\begin{figure}
\includegraphics[width=3.2in, height=1.85in,clip=]{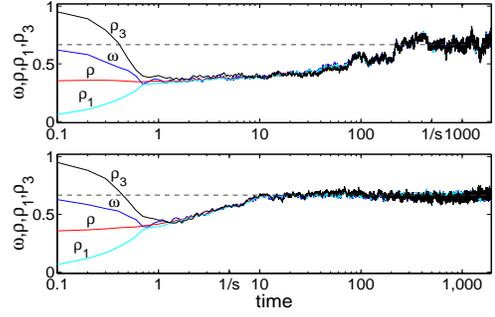}
\caption{{\it (Color online)}.  Timescale separation in the evolution of
the densities $\rho$, $\omega$, $\rho_1$, and
$\rho_3$ on a scale-free network with $\nu=3$ for a  SG with $a=d=1,b=9, c=5$ and $N=10^4$, see text.
Numerical results for typical
 single-realization trajectories for  $s=0.002$ (top) and $s=0.2$ (bottom).
In both panels, initially
$\rho_{k> \mu_1}(0)=1,\rho_{k\leq \mu_1}(0)=0$.
As  eye guides,
 the  dashed line $\rho_*=2/3$ and the times $t=s^{-1}$ are shown.}
 \label{SDnets_fig1}
\end{figure}
\begin{eqnarray}
\label{T_IP}
T^{+} (\rho_k)\!&\equiv\!&T^+_k=(n_k/\mu_1)\left[1+s(b-d)(1-\rho)\right]k(1-\rho_k) \omega\nonumber\\
T^{-} (\rho_k)\!&\equiv\!&T^-_k=(n_k/\mu_1)\left[1-s(a-c)\rho\right]k \rho_k(1-\omega).
\end{eqnarray}
We notice that $T^{\pm}_k$ are nonzero provided that the mean degree $\mu_1$ does not diverge with $N\to \infty$~\cite{nu2}.
In the limit of weak selection intensity ($s\ll 1$), one can use the diffusion theory to treat the birth-death process defined
 by (\ref{T_IP})~\cite{MA10}. This yields a multivariate backward Fokker-Planck equation (FPE) whose generator reads
\begin{equation}\label{fullGen}
{\cal G}(\{\rho_k\})\!=\!\sum_k \left[\frac{(T^+_k\!-\!T^-_k)}{n_k} \frac{\partial}{\partial \rho_k} + \frac{(T^+_k\!+\!T^-_k)}{2Nn_k^2}\frac{\partial^2}{\partial \rho_k^2} \right],
\end{equation}
with time increments
$\delta t=N^{-1}$~\cite{Gardiner,Kimura}.
Furthermore, in the weak selection limit ($s\ll 1$),
the analysis can be simplified using a  timescale separation~\cite{VotNets,VotNets2} (see also \cite{EGTgraphs-bis}).
When $t\ll s^{-1}$ the selection pressure is negligible and  $\rho$ is conserved~\cite{VotNets3}. In fact, using
(\ref{T_IP}) at mean field level gives $(d/dt)\bar{\rho}=s(a-b-c+d) \bar{\omega}(1-\bar{\omega})(\bar{\rho}-\rho_*)$ [the upper
bar denotes the ensemble average]. This indicates that
$\bar{\rho}$ relaxes to its metastable value $\rho_*$ on a timescale $t \sim s^{-1}\gg 1$,
see Fig.~\ref{SDnets_fig1}. At mean field level,
Eqs.~(\ref{T_IP})  also yield
 $(d/dt)\bar{\rho}_k=(T^+_k(\bar{\rho}_k)-T^-_k(\bar{\rho}_k))/n_k=(k/\mu_1)$
$\times\left\{\bar{\omega}-\bar{\rho}_k+s[(b\!-\!d)\bar{\omega}(1\!-\!\bar{\rho})(1\!-\!\bar{\rho}_k)+(a\!-
\!c)(1\!-\!\bar{\omega})\bar{\rho}_k\bar{\rho}]\right\}$. This indicates that after a timescale
of order ${\cal O}(1)$,  $\bar{\rho}_k\approx \bar{\omega}$, and also $\bar{\rho}\approx\bar{\omega}$
since $\bar{\rho}=\sum_k \bar{\rho}_k n_k$. With $\bar{\rho}_k\approx\bar{\omega}\approx\bar{\rho}$,
 the rate equation for $\bar{\rho}_k$ becomes $(d/dt)\bar{\rho}_k\simeq -(k/\mu_1)(b-d)s(1-\bar{\rho}_k)\bar{\rho}_k(\bar{\rho}_k/\rho_*-1)$.
Hence, while after a time of order ${\cal O}(1)$,  $\bar{\rho}_k\approx \bar{\omega} \approx \bar{\rho}$,
all these quantities slowly approach $\rho_*$
after a time $t\sim s^{-1}$. This is illustrated
in Fig.~\ref{SDnets_fig1} where all trajectories rapidly coincide
 and then attain  $\rho_*$ when  $t\sim s^{-1}$.
As fixation occurs on much longer timescales than $s^{-1}$, we  approximate
that on average $\rho_k\approx\rho\approx \omega$ in the same vein as in \cite{VotNets}. With the definition of $\omega$, yielding
$\partial_{\rho_k}\to (k n_k/\mu_1)\partial_{\omega}$, and the change of variables
$\rho_k \to \omega$, Eq.~(\ref{fullGen}) becomes the  {\it effective single-coordinate}
FPE generator:
\begin{eqnarray}
 \label{bFPop_ap}
{\cal G}_{\rm eff}(\omega)=
\frac{\omega (1-\omega)}{N_{{\rm eff}}}\left[-\sigma (\omega - \rho_*)\frac{\partial}{\partial \omega}
+\frac{1}{2}\frac{\partial^2}{\partial \omega^2}
\right].
\end{eqnarray}
The drift term is  proportional to  $\sigma\equiv 2(b-d)N_{{\rm eff}} s_{{\rm eff}}/\rho_*$, where the effective population size and selection
intensity are
 $ N_{{\rm eff}}\equiv N~(\mu_1)^3/\mu_3\; \text{and}\; s_{{\rm eff}}\equiv s~\mu_2/(\mu_1)^2$.
For scale-free networks  with  degree distribution
$n_k \propto k^{-\nu}$ and finite average degree (i.e. $\nu>2$)~\cite{nu2},
the maximum degree is $k_{max}\sim N^{1/(\nu-1)}$~\cite{KS02}.
We thus obtain
 the moments
  $\mu_m$~\cite{VotNets} that yield the scaling of $\sigma$ and $\sigma_{\rm re}\equiv N_{\rm eff} s_{\rm eff}$:
\begin{eqnarray}
 \label{Neff}
\sigma \propto
\sigma_{\rm re}=sN\,\frac{\mu_1\mu_2}{\mu_3}
\sim\begin{cases}
sN,  & \nu> 4\\
sN/\ln{N}, & \nu=4\\
sN^{(2\nu-5)/(\nu-1)}, & 3<\nu<4\\
s\sqrt{N}\ln{N}, & \nu=3\\
sN^{(\nu-2)/(\nu-1)}, & 2<\nu<3.
\end{cases}
\end{eqnarray}
To understand this nontrivial scaling, we focus on scale-free graphs
with $2 < \nu < 3$  characterized by the divergence of $\mu_2$ and $\mu_3$ (when $N\to \infty$).
 Such networks comprise nodes of high degree (hubs) causing the reduction of the
 system's effective size, $N_{\rm eff}\sim N^{(2\nu-5)/(\nu-1)}\ll N$,
and of the system's relaxation time $t\sim s_{\rm eff}^{-1}$ to $\rho_*$, with  $s_{\rm eff}\sim sN^{(3-\nu)/(\nu-1)}\gg s$.
As a result, the fluctuations intensity ($\propto N_{\rm eff}^{-1/2}$) and the drift strength ($\propto s_{\rm eff}$) are
 both enhanced by the topology. Yet, their product
$N_{\rm eff} s_{\rm eff}\sim sN^{(\nu-2)/(\nu-1)}\ll Ns$ indicates that their combined effect
drastically reduces the MFT (see below). We have also checked that our effective theory (\ref{bFPop_ap})
 is applicable  when
 $s_{\rm eff}^2\ll N_{\rm eff}^{-1}$,
 i.e. over a broader range of $s$ than on complete graphs when $2<\nu<4$~\cite{validFPE}.

{\it Fixation properties.}
\begin{figure}
\includegraphics[width=3.2in, height=1.9in,clip=]{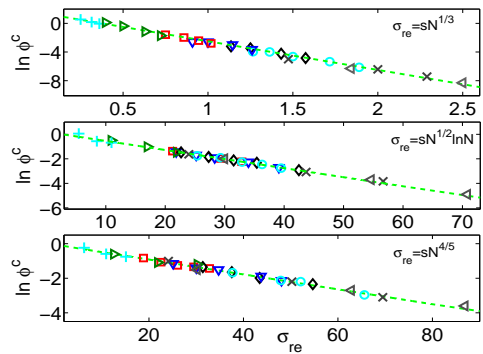}
\caption{{\it (Color online)}.
Probability $\phi^C$ versus  $\sigma_{\rm re}$
for SGs with $a=d=1$, $b=1.05$, $c=1.075$
 and $s=0.025$ ($+$), $0.05$ $(\triangleright)$, $0.075$ $(\square)$, $0.1$ $(\nabla)$, $0.125$
$(\diamond)$, $0.15$ $(\circ)$, $0.2$ $(\times)$, $0.25$ $(\triangleleft)$.
Numerical results for $\nu=2.5$ (top),
$\nu=3$ (middle),  $\nu=3.5$ (bottom) collapse along the straight dashed lines obtained from (\ref{FixProb}), see text.
Here  $N=400 - 4000$ and
initially  $\rho_k=\rho=\omega=100/N$. Error bars are of
size of the symbols.}\label{SDnets_fig2}
\end{figure}
 Evolutionary dynamics is characterized by the fixation probability $\phi^C(\omega)$
that a system with initial degree-weighted density
$\omega$ is taken over by cooperators.
In the framework of the effective diffusion theory and using~(\ref{bFPop_ap}) the fixation probability obeys
${\cal G}_{\rm eff}(\omega)\phi^C(\omega)=0$  with
 boundary conditions (BCs)
$\phi^C(0)=1-\phi^C(1)=0$~\cite{weaksel,Gardiner}. The  solution reads
\begin{equation}
 \label{FixProb}
\phi^C(\omega)=\frac{
{\rm erfi}\left[\rho_*\sqrt{\sigma}\,\right] -
{\rm erfi}\left[(\rho_*-\omega)\sqrt{\sigma}\,\right]
}{{\rm erfi}\left[\rho_*\sqrt{\sigma}\,\right]+{\rm erfi}\left[(1 -\rho_*)\sqrt{\sigma}\,\right]},
\end{equation}
where ${\rm erfi}(z)\equiv\frac{2}{\sqrt{\pi}}\int_{0}^z e^{u^2} du$.
Let us consider the (biologically relevant) case of a small initial density of cooperators such
 that $\omega \ll 1$, weak selection~\cite{Kimura,weaksel},
and a large population such that $\rho_*^2 \sigma \gg 1$ and metastability is guaranteed.
Using the asymptote ${\rm erfi}(x)\sim e^{x^2}$ for $x\gg 1$ in Eq.~(\ref{FixProb}),
we distinguish two cases: (i) when $\rho_*<1/2$, $\ln \phi^C\simeq -(1-2\rho_*)\sigma$;
 (ii) when $\rho_*>1/2$ and $\omega>2\rho_*-1$, $\ln (1-\phi^C)\simeq -(2\rho_*-1)\sigma$,
 while $\ln (1-\phi^C)\simeq -\omega(2\rho_*-\omega)\sigma$
if $\rho_*>1/2$ and $\omega<2\rho_*-1$.
In Fig.~\ref{SDnets_fig2} (and Fig.~\ref{SDnets_fig3}),
for each value of $s$ the numerical results have been rescaled by a constant
to test the scaling (\ref{Neff}). The linear data collapse
and stretched exponential dependence $\ln \phi^C\sim -sN\mu_1\mu_2/\mu_3$ predicted by
 (\ref{Neff},\ref{FixProb}) is indeed clearly observed in Fig.~\ref{SDnets_fig2}. Since $\ln \phi^C\sim -sN$ on complete graphs~\cite{MA10}, this  demonstrates how
the scale-free structure drastically affects
 the fixation probability.

Another quantity of great interest is the (unconditional) MFT $\tau(\omega)$ -- the mean time necessary to reach
an absorbing boundary. Here, using Eq.~(\ref{bFPop_ap}) the  MFT is obtained by solving ${\cal G}_{\rm eff}(\omega)\tau(\omega)=-1$
with  BCs
$\tau(0)\!=\!\tau(1)\!=\!0$~\cite{Gardiner}. Using standard methods~\cite{Kimura,Gardiner}, we obtain  
$\tau(\omega)=2 N_{\rm eff}\left[(1-\phi^C(\omega))\!\int_0^\omega dy \frac{e^{-\Theta(y)}}{y(1\!-\!y)}\! \int_{0}^{y} dz\,e^{\Theta(z)}\right.\nonumber\\
+ \left.\phi^C(\omega)\!\int_\omega^1 dy \frac{e^{-\Theta(y)}}{y(1\!-\!y)}\! \int_{y}^{1} dz\,e^{\Theta(z)}\right];
\;\Theta(z)\equiv \sigma z(z\!-\!2\rho_*).$
For $\rho_*^2 \sigma\gg 1$ the inner integrals can be computed by
expanding $\Theta(z)$ around its extremal values ($z=0$ for $z\in [0,\rho_*]$ and  $z=1$ for $z\in [\rho_*,1]$),
while the outer integral is computed via the saddle-point approximation around $\omega=\rho_*$. To leading order, one thus obtains a stretched exponential dependence on $N$: $\tau(\omega)\sim (1-\phi^C(\omega)) e^{\sigma\rho_*^2}$ when $\omega > \rho_*$ and $\tau(\omega)\sim \phi^C(\omega) e^{\sigma(1-\rho_*)^2}$  otherwise.
For example,
when $\rho_*<1/2$  this  gives
 (see Fig.~\ref{SDnets_fig3})
\begin{equation}
 \label{MFTsmall}
\ln{\tau(\omega)}\simeq \sigma\rho_*^2\propto \sigma_{\rm re}.
\end{equation}

When the initial number of cooperators is not too low, the long-lived metastable state
is entered prior to fixation and the  MFT~(\ref{MFTsmall}) is independent of the initial condition~\cite{AM,MA10}.
Eq.~(\ref{MFTsmall}), confirmed by Fig.~\ref{SDnets_fig3},
implies that for scale-free networks with $2<\nu<4$ fixation occurs much more rapidly
than on complete graphs, a phenomenon called ``hyperfixation''
in  genetics~\cite{FixNet}.
\begin{figure}
\includegraphics[width=3.4in, height=1.9in,clip=]{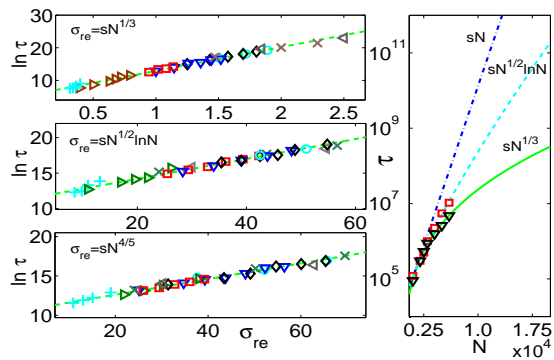}
\caption{{\it (Color online)}.
(Left) $\tau$ vs  $\sigma_{\rm re}$ for $\nu=2.5$ (top), $\nu=3$
(middle), and $\nu=3.5$ (bottom).
 Numerical results (symbols) collapse along lines (dashed) in agreement with (\ref{MFTsmall}), see text.
Symbols and parameters are as in Fig.~\ref{SDnets_fig2}. (Right)
$\tau$ vs  $N$ on semi-log scale with $s=0.1$: numerical results for $\nu=2.5$~($\nabla$) and $\nu=3$
($\square$) agree with (\ref{MFTsmall}), shown as solid/dashed lines.
The result on  complete graphs  ($\ln{\tau}\sim sN$) is sketched as eye guide (dash-dotted).
 In all panels, initially $\rho_k=\rho=0.5$.
}
 \label{SDnets_fig3}
\end{figure}

{\it Discussion \& conclusion.}
We have studied metastability and fixation of evolutionary processes on scale-free networks
in the realm of EGT. For the sake of concreteness, we have
focused on ``snowdrift games'' evolving with the LD~\cite{VotNets}
and characterized by a long-lived (metastable) coexistence state.
The evolutionary dynamics has been described by a birth-death process
from which we have derived an effective diffusion theory by exploiting
  a timescale separation occurring at weak selection intensity.
The  probability and mean fixation time (MFT) have been computed
from the corresponding backward Fokker-Planck equation.
These quantities exhibit a stretched-exponential dependence on the population size, in
stark contrast with their non-spatial counterparts. We have checked with various update rules that
the stretched-exponential behavior is
 a generic feature of metastability on scale-free graphs
that also characterizes the fixation probability of coordination games~\cite{inprep,EGT}.
Here, important  consequences of the
stretched-exponential behavior are a drastic reduction of the MFT and the
possible enhancement of the fixation probability of a few mutants with respect to a non-spatial setting.
These anomalous fixation and metastability properties reflect the strong influence of the network's structure on
 evolutionary processes.


\begin{thebibliography}{99}
\bibitem{dynamics}
One can also use the voter model  or invasion process update rules. These and the LD
are equivalent on regular graphs, but markedly different on
scale-free graphs~\cite{VotNets}.
\bibitem{nu2}
Our model does not apply to  (unrealistic) graphs
with $\nu\leq 2$, i.e. with diverging
 average degree $\mu_1$, see Eqs.~(\ref{T_IP}).
\bibitem{validFPE}
E.g.,  (\ref{bFPop_ap}) is valid for  $s^2\ll  N^{-1/(\nu-1)}$ when $2<\nu<3$, while
the condition on complete graphs is $s^2\ll N^{-1}$~\cite{MA10}.
\end{thebibliography}
\end{document}